\documentclass[12pt,a4paper]{article}
\usepackage[left=18mm, top=10mm, right=18mm, bottom=20mm, nohead, foot=10mm]{geometry}

\usepackage[affil-it]{authblk}

\usepackage{graphicx}
\usepackage{dcolumn}
\usepackage{bm}
\usepackage{color}
\usepackage{url}
\usepackage{tabularx}
\usepackage{array}
\usepackage{booktabs}
\usepackage{comment}
\usepackage{hyperref}
\usepackage{footnote}
\usepackage[normalem]{ulem}

\date{\today}
\begin{document}

\title{\Large \textbf{New spiral state and skyrmion lattice in 3D model of chiral magnets}}
\author[1]{\large F.\,N. Rybakov\thanks{f.n.rybakov@gmail.com}}
\author[1]{\large A.\,B. Borisov}
\author[2]{\large S. Bl\"ugel}
\author[2]{\large N.\,S. Kiselev\thanks{n.kiselev@fz-juelich.de}}
\affil[1]{\normalsize M.\,N. Miheev Institute of Metal Physics of Ural Branch of Russian Academy of Sciences, Ekaterinburg 620990, Russia}
\affil[2]{\normalsize Peter Gr\"unberg Institut and  Institute for Advanced Simulation, Forschungszentrum J\"ulich and JARA, D-52425  J\"ulich, Germany}
\renewcommand\Authands{ and }
\date{}

\maketitle

\begin{abstract}
We present the phase diagram of magnetic states  for films of isotropic chiral magnets calculated as function of applied magnetic field and thickness of the film.
We have found a novel magnetic state driven by the natural confinement of the crystal, localized at the surface and stacked on top of the conical bulk phase.
This magnetic surface state has a three-dimensional (3D) chiral spin-texture described by the superposition of helical and cycloidal spin spirals.
This surface state exists for a large range of applied magnetic fields and for any film thickness beyond a critical one.    
We also identified the whole thickness and field range for which the skyrmion lattice becomes the ground state of the system.
Below a certain critical thickness the surface state and bulk conical phase are suppressed in favor of the skyrmion lattice.   
Unraveling of those phases and the construction of the phase diagram became possible using advanced computational techniques for direct energy minimization applied to a basic 3D model for chiral magnets.
Presented results provide a comprehensive theoretical description for those effects already observed in experiments on thin films of chiral magnets, predict new effects important for applications and open perspectives for experimental studies of such systems.
\end{abstract}

\section{Introduction}%-----------------------------
\begin{figure*}[ht]
\centering
\includegraphics[width=1.0\columnwidth]{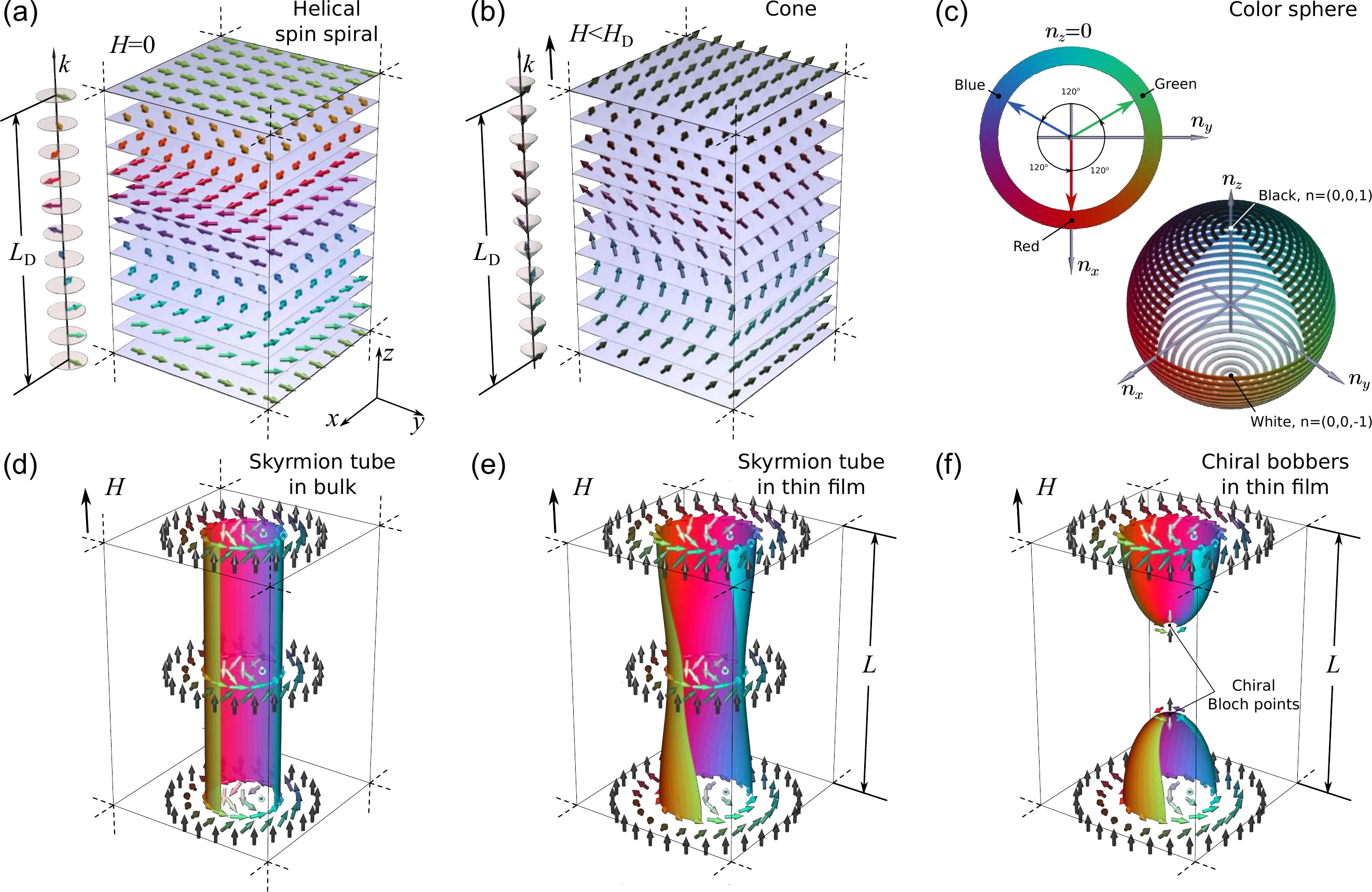}
\caption{ 
\textbf{Schematic representation of different modulated states in bulk and thin films of chiral magnets.} 
(a) Helical spin spiral at zero field with wave vector $\mathbf{k}$ pointing along the $z-$axis.
(b) Conical state with inclined magnetization and wave vector along the  magnetic field.
(c) Color sphere illustrating the notation for the component of the vector field used in all figures, here and below.
(d)--(f) The vector field and cross sections of corresponding isosurfaces, $n_z=0$, for: skyrmion tube with homogeneous magnetization along the radial symmetry axis (d), for skyrmion tube in thin film of chiral magnet with inhomogeneous magnetization in all three spatial directions and the twist induced by the free surface (e), chiral bobbers --  hybrid particlelike states localized near the surface of the chiral magnet characterized by smooth magnetization distribution and presence of singularity -- chiral Bloch point at finite distance from the surface (f), for details see Ref.~\cite{Rybakov_15}. Note, the section of the isosurfaces in (d)-(f) corresponds to $n_z>0$, $n_x<0$.
}
 \label{Intro}
\end{figure*}

Chiral magnets (ChM) are a distinct class of magnetic crystals. Contrary to the classical ferro- and antiferromagnets the ground state of ChM is an incommensurate homochiral spin spiral -- the  spiral with an unique sense of magnetization rotation, see figure~\ref{Intro}(a). 
The interaction which is responsible for the stabilization of such spin spirals is an antisymmetric exchange, also known as Dzyaloshinskii-Moriya interaction (DMI)\cite{Dzyaloshinskii, Moriya}, which appears in crystals with  broken inversion symmetry.
For instance, to this class of magnetic materials belong different Si- and Ge-based alloys such as MnSi \cite{Yu_15} and FeGe \cite{Yu_11}, Mn$_{1-x}$Fe$_x$Ge \cite{Shibata_13},  Mn$_{1-x}$Fe$_x$Si \cite{Yokouchi_14}, Fe$_{1-x}$Co$_x$Si \cite{Yu_10}.
Such alloys with B20 crystal symmetry can be referred to as the distinct class of so-called isotropic ChM (IChM). Such a classification reflects the dominant role of DMI and Heisenberg exchange, which are assumed to be isotropic in all spatial directions, while in the frame of the basic model, the contribution of magnetocrystalline anisotropy can be neglected.

The special interest to such materials arose after the 
breakthrough results on the direct observation of magnetic skyrmions in thin films of ChM \cite{Yu_10}. 
The experimental discovery of magnetic skyrmions together with the conceptual idea of a revolutionary new type of magnetic memory gave an additional impetus to the research in this field \cite{Kiselev_11,Fert_13}.

Here we present results of our theoretical calculations for the magnetic field induced transitions in thin films of IChM, which allowed us to identify the critical film thickness above which the skyrmions never appear as the ground state of the system.
Realizing that even in the simplest one-dimensional (1D) model of ChM the ground state exhibits periodic modulations \cite{helix}, in the three-dimensional (3D) systems one should expect the appearance and coexistence of complex inhomogeneous phases, which are of major interest for both fundamental research and  practical applications.
Indeed, in this paper we identify an earlier unknown phase, that we term \textit{stacked spin spirals phase} (StSS). The calculated phase diagram exhibits a wide range of existence of this new phase. 
This new phase exhibits periodic modulations in all three spatial directions and can be considered as the coexistence of the conical phase and the complex spin spirals localized near the surfaces.
We found that such a state should appear in a wide range of film thicknesses as well as in the bulk samples.
We provide a comprehensive description of this phase together with our suggestions  how such a state can be experimentally detected.  

It is worth mentioning that the existence of the magnetic skyrmions in magnetic crystals with DMI has been theoretically predicted by Bogdanov in 1989 \cite{Bogdanov_89}, but it took more then twenty years to discover them experimentally.
One of the reason behind is the modification of the conventional energy balance in thin films of ChMs in comparison to the bulk crystals.
Indeed, according to the theory developed by Bogdanov and co-workers, see e.g. Ref.~\cite{Bogdanov_14}, the chiral skyrmion tubes, see figure~\ref{Intro}(d), in bulk crystals of ChMs with relatively weak magnetocrystalline anisotropy may appear only as a metastable state.
Their energy is always higher than the energy of the conical phase, figure~\ref{Intro}(b), which dominates in bulk crystals almost in the whole range of magnetic fields and temperatures.
This result is consistent with many experimental studies on bulk ChM over the past decades, where skyrmions were not observed. 
The only exception is the so-called A-phase -- the high-temperature region in the bulk phase diagram, just below the ordering temperature, where skyrmions has been proposed to exist due to thermal fluctuations \cite{Muelbauer_09}.
However, the real nature of the A-phase still is under debate \cite{Muelbauer_09,Maleev, Wilhelm_11}.
The theoretical models predict that in bulk crystals of ChMs, the magnetic skyrmions can be stabilized due to the strong cubic or uniaxial anisotropy \cite{Bogdanov_14,Bogdanov_08} or special crystal symmetry, which suppress the formation of the conical phase \cite{Bogdanov_89,Bogdanov_94}.
However, such theoretical models as well as thermal fluctuations can not be considered as the main mechanism for skyrmion stabilization in stand-alone thin films of B20-alloys where the skyrmions are observed in a wide range of temperatures, much lower than the ordering temperature.

As has been shown in Ref.~\cite{Rybakov_13} the key to the understanding of the mechanism of skyrmion stabilization in thin films of ChM are presence of the free surfaces and the three-dimensional rather than the two-dimensional structure of the equilibrium skyrmions.
As shown in figure~\ref{Intro}(e), the solution for the skyrmion tube in a thin film is characterized by the twist of magnetization with respect to the normal vector of the film surfaces.
One can compare the twisted iso-surface of the skyrmion tube in figure~\ref{Intro}(e) and the homogeneous non-twisted skyrmion tube in (d). 
The magnetic moments in the top surface layer are slightly turned towards the center of the skyrmion, while in the bottom of the film they are slightly turned outwards the center.
The spin structure on the top and bottom surfaces corresponds to a certain intermediate configuration in between of pure Bloch- and Neel-type of skyrmions. 
Such surface induced twist propagates from one surface to another through the whole film.
Note, it mainly affects the magnetic spins near the surfaces where spins are weakly coupled to each other because of reduced number of neighbors at the free surface.
Far from the film surface, the spin structure of such a 3D skyrmion remains almost the same as in a homogeneous skyrmion tube.

The energy gain by the DMI contribution accumulated along the film thickness reduces the total energy of the state such that within a certain range of magnetic fields and film thicknesses the skyrmion tube becomes energetically more favorable than the conical phase.
Moreover, it has been shown earlier that the same effect is responsible for the stability of chiral bobber -- particlelike objects localized in all three spatial directions near the surface of the chiral magnet \cite{Rybakov_15}, see figure~\ref{Intro}(f).
Note, because of the presence of a singularity, the chiral Bloch point, at low temperatures the chiral bobbers  appear exclusively as a metastable state. 

It is obvious that for very thick films, the relative energy contribution of the surface twist becomes very small, while the main contribution to the energy of skyrmion comes from the volume part of the film. 
Thereby, there should be a critical thickness above which the energy gain of the surface twist is not anymore sufficient to provide enough energy gain to stabilize a skyrmion lattice.
In order to identify the range of thicknesses and magnetic fields defining the range of stability for the skyrmion lattice and other states, we have calculated a phase diagram  for the film of isotropic ChM in a wide range of thicknesses and applied magnetic fields, which is presented in section~\ref{PphaseDiargam}.

\section{Model}\label{model}
The basic model for IChM include three main energy terms: the Heisenberg exchange interaction, the DMI and Zeeman energy term \cite{Bar,Bak}: 
\begin{equation}
\mathcal{E} \!= \mathcal{E}_0+\!\int\limits_{V}  
\!\mathcal{A} \left( {{\partial}_x\mathbf{n}}^2 + {{\partial}_y\mathbf{n}}^2 + {{\partial}_z\mathbf{n}}^2   \right) +  
\mathcal{D}\, \mathbf{n}\cdot [\nabla \times \mathbf{n}   ] + 
H M_\mathrm{s} (1-{n}_z)  \mathrm{d}{\bf r}, 
\label{E_tot_m}
\end{equation} 
where  $\mathbf{n} \equiv \mathbf{n}(\mathbf{r})$ is a continuous unit vector field defined everywhere except at the singular points. $\mathcal{E}_0$ is the energy of the saturated ferromagnetic state. $\mathcal{A}$ and $\mathcal{D}$ are micromagnetic constants for exchange and DMI, respectively, and $M_\mathrm{s}$ is the magnetization of the material -- the total magnetic dipole moment per unit volume. 

We use the continuum model as the most general approach to describe long-period incommensurate magnetic structures. 
The results presented here can be easily generalized for a wide class of the systems. 
The functional (\ref{E_tot_m}) has to be considered as a continuum limit for the classical  spin models  e.g.\ as  simplified models considering a simple cubic lattice \cite{Han_09}, and advanced models, which take into account the exact B20 crystal symmetry \cite{Chizhikov12,Chizhikov13} .

The lowest period of an incommensurate spin spiral and equilibrium period of the conical phase, $L_D$, as well as the critical field corresponding to the saturation field of the conical phase, $H_\mathrm{D}$, have analytic solutions, which couple the material parameters with experimentally measurable quantities \cite{helix,Bogdanov_94}:
\begin{equation}
L_\mathrm{D}=4\pi\frac{\mathcal{A}}{\mathcal{D}},\ \
H_\mathrm{D}=\frac{\mathcal{D}^2}{2M_\mathrm{s}\mathcal{A}}.
\label{LH_params}
\end{equation}

The comparison of the energy density of each of the equilibrium states obtained by direct energy minimization of the functional (\ref{E_tot_m}) allows one to identify the geometrical and material parameters corresponding to the phase transitions. For details of the energy minimization technique and calculation of the phase diagram, see section~\ref{method}.  

\section{Results}\label{Results}
\subsection{Phase diagram}\label{PphaseDiargam}
\begin{figure*}[ht]
\centering
\includegraphics[width=1.0\columnwidth]{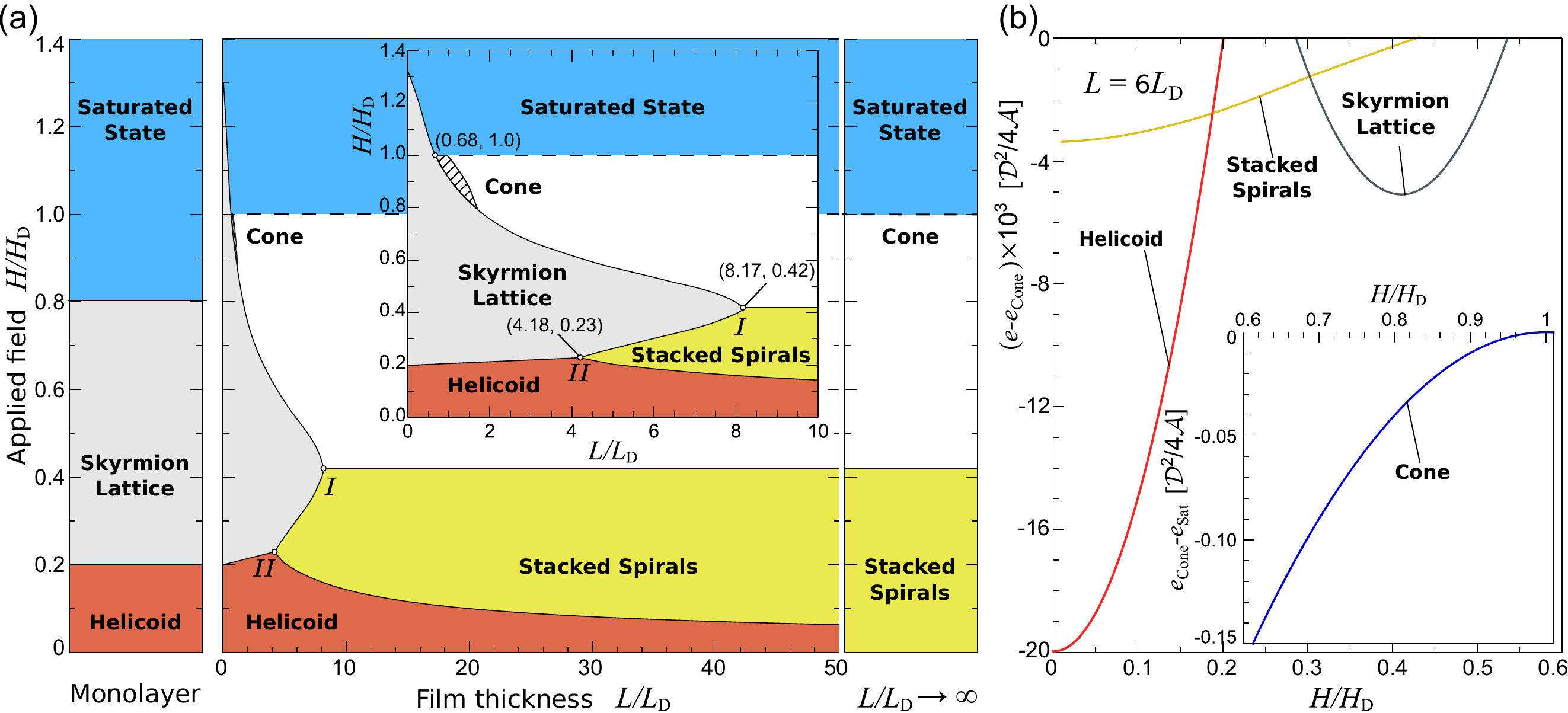}
\caption{ 
\textbf{The phase diagram of magnetic states in film of isotropic chiral magnet.}
(a) The phase diagram of magnetic states in reduced units of film thicknesses $L$ and magnetic field $H$ applied normally to the film surface.
The central panel corresponds to the reduced film thickness varied between 0 and  $L=50L_\mathrm{D}$ ($L_\mathrm{D}$ is the period of helicoid at $H$=0). Inset shows details of the phase diagram for thin films, $0<L\le 10L_\mathrm{D}$.   
Note, everywhere within the range of stability of the conical phase, the isolated chiral bobber corresponds to the lowest metastable state, the only exception is the small dashed area, where isolated skyrmions within the conical phase becomes the metastable state with the lowest energy.
The left panel corresponds to the case of monolayer with surface induced DMI.
The right panel corresponds to the case of infinitely thick film, which qualitatively corresponds to the bulk crystal.        
Small open circles marked as $I$ and $II$, with corresponding coordinates indicated as $(L/L_\mathrm{D}$, $H/H_\mathrm{D})$ correspond to the triple-points.
(b) The average energy density of the  different phases as a function of applied field for the film thickness, $L=6L_\mathrm{D}$. 
}
\label{PD}
\end{figure*}

In the phase diagram presented in figure~\ref{PD}(a) the thickness of the film, $L$, and magnetic field $H$ are given in reduce units, where $L_\mathrm{D}$ and $H_\mathrm{D}$ are the functions of material parameters $\mathcal{A}$, $\mathcal{D}$ and $M_\mathrm{s}$, see Eq.~(\ref{LH_params}).
The unique pair of parameters $L_\mathrm{D}$ and $H_\mathrm{D}$ can be considered as a \textit{fingerprint} of each particular IChM. They can be measured experimentally, which allows to rescale the phase diagram, figure~\ref{PD}(a), in real units of film thickness, $L$ and magnetic field $H$. 

The solid lines in figure~\ref{PD}(a)  correspond to the first-order phase transitions between helical spin spiral (red), skyrmion lattice (grey), conical phase (white) and new up-to-now unknown phase, which we call \textit{stacked spin-spirals state} (yellow) and is discussed in detail in  section~\ref{StSS}.
The horizontal dashed line indicates the second order phase transition between conical and saturated ferromagnetic state (blue).          

The phase diagram for a thin layer of IChM for relatively narrow range of thicknesses, $0\!\!<\!\!L/L_\mathrm{D}\!\!\le\!\!4$, had been  presented earlier in Ref.~\cite{Rybakov_15} . An attempt to generalize such a phase diagram for a wider range of thicknesses, $1\!\!\le \!\!L/L_\mathrm{D}\!\! < \!\! \infty$, had been announced recently \cite{Leonov_15arXiv}.  
The phase diagram presented in Ref.~\cite{Leonov_15arXiv} contains four distinct states and one triple-point.
We have tested these data and got qualitatively and quantitatively different results.
The phase diagram presented in this contribution contains five distinct states and two triple points.

The triple point $I$ defines the critical thickness of the film, $L^*=8.17 L_\mathrm{D}$, above which the skyrmions may exist only as a metastable state. For instance, for MnSi ($L_\mathrm{D}=18$ nm) and FeGe ($L_\mathrm{D}=70$ nm) it gives $L^*_\mathrm{MnSi}\approx150$~nm and $L^*_\mathrm{FeGe}\approx570$~nm.
With decreasing thickness the range of existence for the skyrmion lattice in an fapplied magnetic field becomes wider.
This fact reflects the relative contribution of the surface induced twists, which increases with decreasing thickness.
There is another critical point, for $L/L_\mathrm{D}<0.68$, the conical phase is totally suppressed and becomes energetically unfavorable in the whole range of fields. 

It is important to mention that the chiral surface twist discussed above also introduces an additional modulation in the helical spiral state.
Note, the $\mathbf{k}$-vector of such helical spiral lies in the plane of the film, orthogonal to the applied field.
The surface induced modulation reduces the energy of the helix and in a certain range of fields makes it energetically more favorable than the conical state.
Such a behavior of the system is totally different to the one of the bulk crystals, where theoretically any infinitesimal magnetic field leads to convergence of the helix to the conical phase or more precisely to the StSS according to results presented here, see right panel in figure~\ref{PD}(a). 

Finally, we have to emphasize that the effect of the chiral surface twist is not restricted to the film surfaces, but also appears on the side edges of the sample.
The presence of the edge twist effect has been reported earlier in Ref.~\cite{Surf_twist1} and  has been confirmed recently by directly observation with the Lorentz Transmission Electron Microscopy (TEM) \cite{Du_15}. 
  
It is worth to emphasize that, the continuum 3D model of IChM strictly speaking does not converge to a simple 2D model even in the limiting case of $L/L_\mathrm{D}\rightarrow0$.
In order to illustrate such a discrepancy we added the left panel in figure~\ref{PD}(a), which corresponds to the phase diagram of 2D IChM valid for the single isotropic monolayer or multilayer with interface induced DMI and for ChM of particular crystal symmetry e.g.\ $C_{nv}$, $D_{2d}$ and $S_4$ \cite{Bogdanov_89,Bogdanov_94}. 

\subsection{Stacked spin spirals}\label{StSS}
\begin{figure*}[ht]
\centering
\includegraphics[width=1.0\columnwidth]{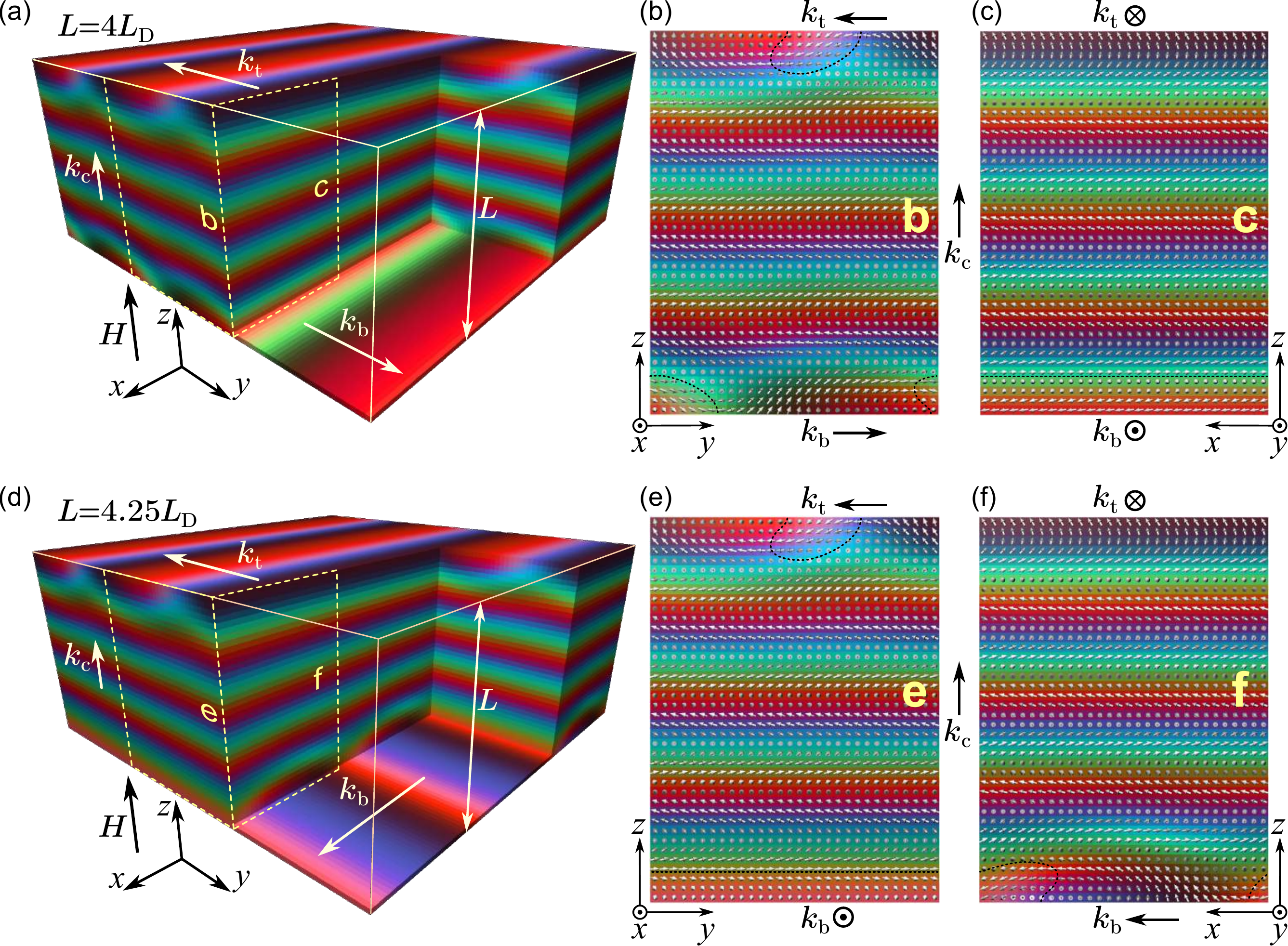}
\caption{ 
\textbf{Stacked spirals states.}
Magnetization distribution in  films of chiral magnets with the local magnetization direction as indicated by the colors according to the code in figure~\ref{Intro}(c). 
(a)--(c) The magnetization distributions for the stacked spin spirals in the film with the thickness of $L=4L_\mathrm{D}$.
$\textbf{k}_\mathrm{c}$ indicates the propagation direction of the conical spiral in the bulk of the film, while $\textbf{k}_\mathrm{t}$ and $\textbf{k}_\mathrm{b}$ indicate the propagation directions of the complex spin spirals on the top and bottom surfaces, respectively. Note, for this thickness  $\textbf{k}_\mathrm{t}$ and $\textbf{k}_\mathrm{b}$ are anti-parallel and both are orthogonal to $\textbf{k}_\mathrm{c}$.
(b) and (c) magnetization distribution in the projections on the cross-sections of the $yz$-- and $xz$--plane, see corresponding dotted rectangles in (a). 
(d)--(f) The magnetization distribution in the film with the thickness of $L=4.25L_\mathrm{D}$. Note, for this thickness the equilibrium state of stacked spirals characterized by the mutually orthogonal vectors $\textbf{k}_\mathrm{t}$, $\textbf{k}_\mathrm{b}$ and $\textbf{k}_\mathrm{c}$. For details of complex surface spin spiral see figure~\ref{Twist-surface}. 
The black dashed curves in (b), (c), (e) and (f) indicate the small volume near the surface with the negative component of magnetization, $n_z<0$, pointing away from the applied field, $\mathbf{H}||z$-axis.
}
\label{Twist}
\end{figure*}

\begin{figure}[ht]
\centering
\includegraphics[width=13cm]{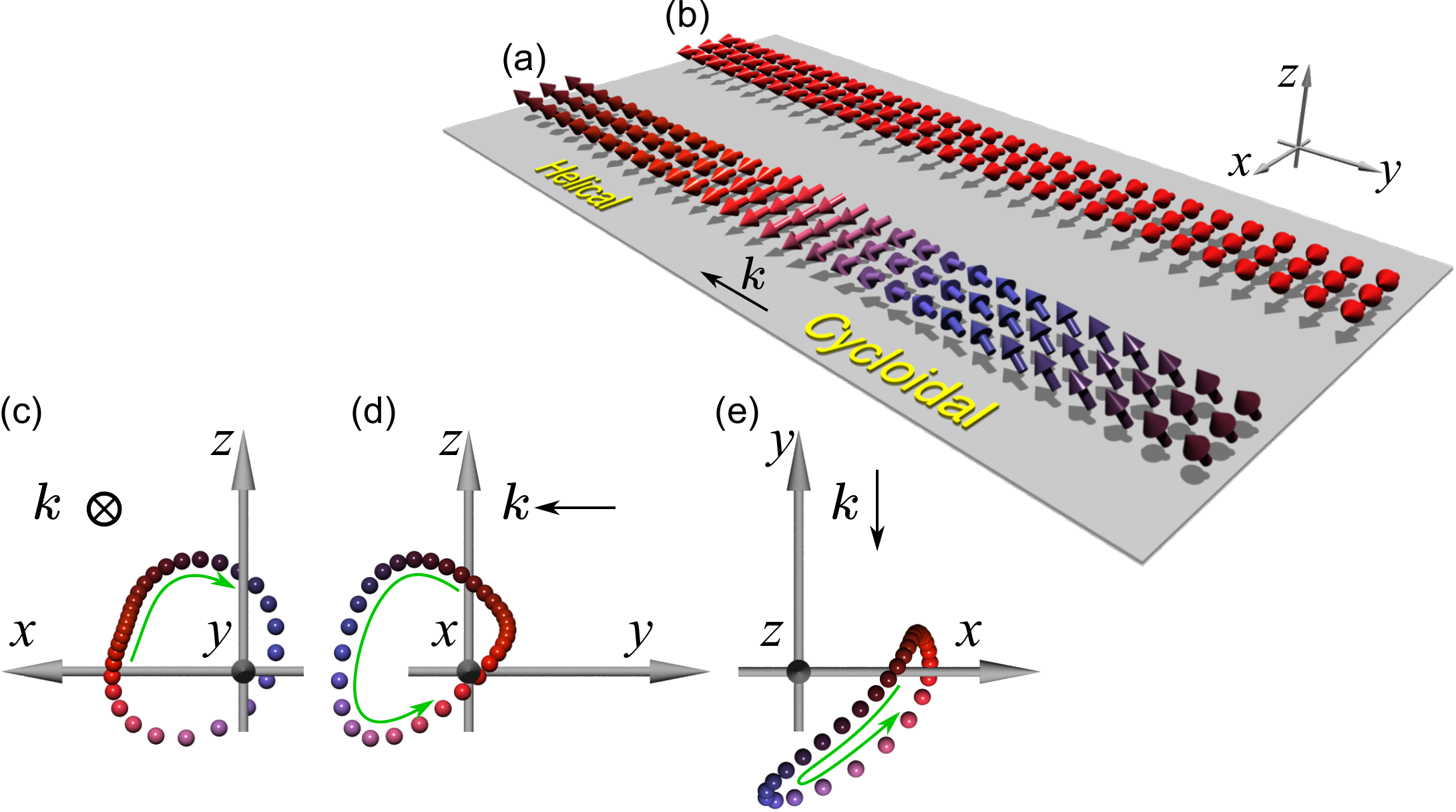}
\caption{
\textbf{Complex spin spiral on the surface of chiral magnet.}
(a) The vector field of the spin spiral with mixed helical and cycloidal modulation of magnetization along the direction propagation  vector $\textbf{k}$.
(b) The homogeneous vector field in the surface layer for the conical phase is added for comparison.
(c)--(e) The trajectories of the spins in the complex spin spiral on the surface of a unit sphere shown in projections on the $xz$--, $yz$-- and $xy$--plane, respectively. 
The green arrows indicate the sense of rotation of the spins. The $\mathbf{k}$-vector indicates the propagation direction of the spiral.
}
\label{Twist-surface}
\end{figure}

\begin{figure*}[ht]
\centering
\includegraphics[width=13cm]{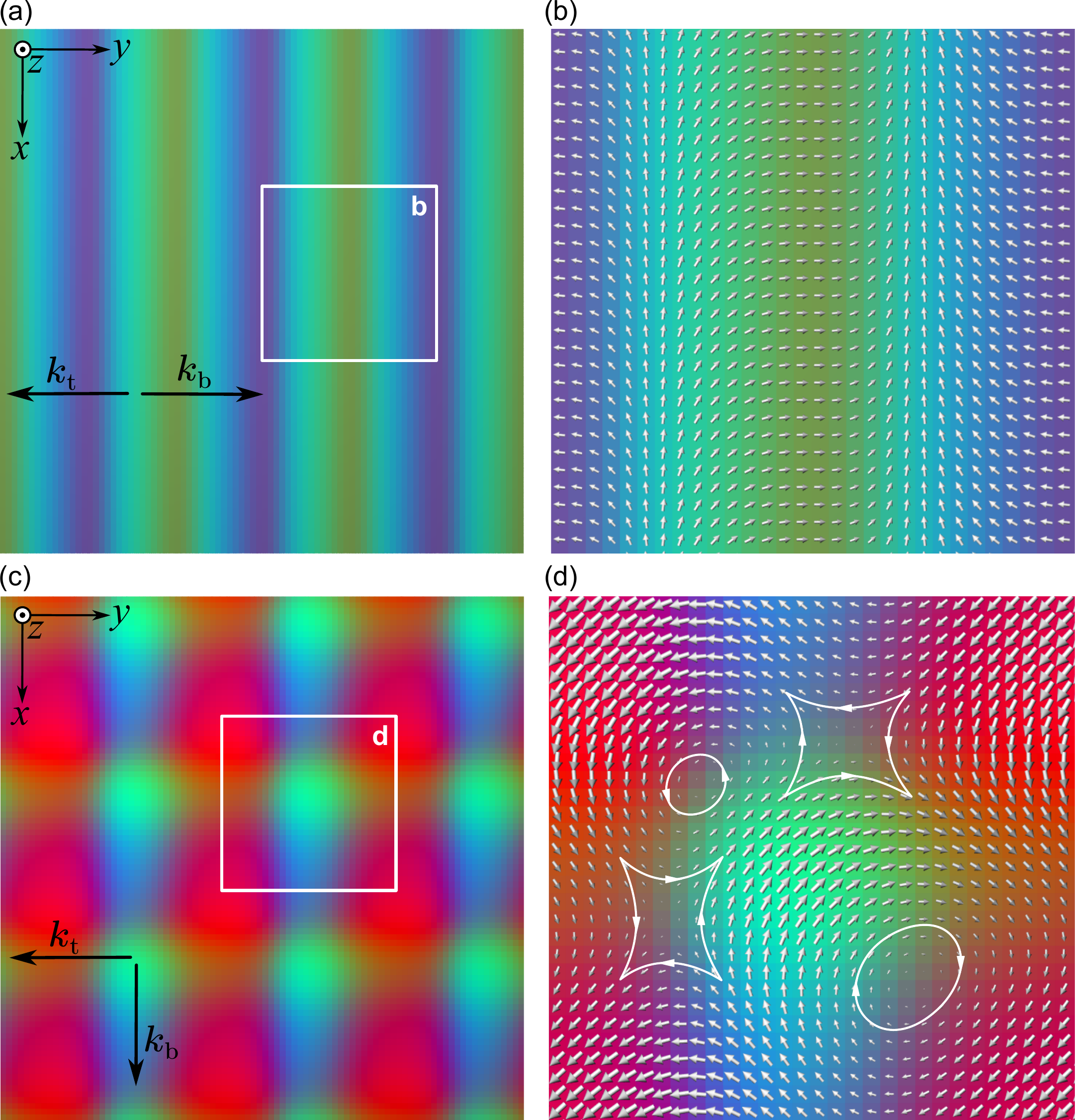}
\caption{ 
\textbf{The magnetization distribution averaged over the film thickness in the stacked spirals states.}
(a) Color map for the averaged magnetization for the film with $L=4L_\mathrm{D}$, see figure~\ref{Twist}(a).
(b) Zoomed view for the corresponding part in (a) with arrows indicating in-plane components of the magnetization, $n_{xy}$.
(c) The averaged magnetization for the film with $L=4.25L_\mathrm{D}$, see figure~\ref{Twist}(d).
(d) Zoomed view for the corresponding part in (c).
The closed white line indicate vortex- and anti-vortex-like part of the average in-plane component of the magnetization.
Note, in each point, the out-of-plane component of the average magnetization is pointing along the applied field $H\parallel z$-axis, $n_z>0$. For better visualization the in-plane components, $n_{xy}$, are magnified 6 times. 
}
\label{TwistAv}
\end{figure*}

A wide range of the phase diagram is occupied by the newly found stacked spin spirals state (StSS).
The triple point $II$ defines the limiting thickness above which the StSS may appear as the global energy minimum.

Figure~\ref{Twist} illustrates the complex spin structure of the StSS obtained by direct minimization of the functional (\ref{E_tot_m}).
The StSS represents the coexistence of the conical phase in the bulk of the sample and the quasi-helical modulation of magnetization localized in the vicinity of the surface of the sample. 
Such free surface induced modulations have finite penetration depth and appear on both the top and bottom surfaces. It exhibits a mixed helical- and cycloidal-like modulation as shown in figure~\ref{Twist-surface}(a).
The clock-wise rotation of magnetization in the helical-like part is chosen as direction of the wave vector $\mathbf{k}$ for such a complex spin spiral, see figure~\ref{Twist-surface}(c).

We found that the relative orientation of the propagation vectors of the spin spirals on the top and bottom surfaces, $\mathbf{k}_\mathrm{t}$ and $\mathbf{k}_\mathrm{b}$, respectively, is thickness dependent.
In other words, the angle $\beta_\mathrm{tb}$ between $\mathbf{k}_\mathrm{t}$ and $\mathbf{k}_\mathrm{b}$ for the equilibrium StSS varies as function of the film thickness.
In particular, under the condition $L=(n+\frac12)L_\mathrm{D}$, where $n$ is an integer number, the equilibrium $\beta_\mathrm{tb}=0$ and gradually varies with the thickness: for $L=(n+\frac34)L_\mathrm{D}$, $\beta_\mathrm{tb}=90^{\circ}$, for $L=(n+1)L_\mathrm{D}$, $\beta_\mathrm{tb}=180^{\circ}$, and for $L=(n+1\frac14)L_\mathrm{D}$, $\beta_\mathrm{tb}=90^{\circ}$.
Figure~\ref{Twist}(a) illustrates the case of $\beta_\mathrm{tb}=180^{\circ}$, $L=4L_\mathrm{D}$, while (d) illustrates the case of $\beta_\mathrm{tb}=90^{\circ}$, $L=4\frac14 L_\mathrm{D}$.
Note, because of the relatively weak energy dependence upon the $\beta_\mathrm{tb}$ in the case of thick films, one should expect a significant influence of the magnetocrystalline anisotropy to the orientation of the vectors $\mathbf{k}_\mathrm{t}$ and $\mathbf{k}_\mathrm{b}$ with respect to certain crystallographic directions.

The modulation of the out-of-plane component of the magnetization of the surface spin spiral should lead to  appearances of the stray fields above the sample, which in turn can be detected by Magnetic Force Microscopy (MFM) in films as well as in bulk crystals. 
However, with ordinary MFM technique it seems to be hard to distinguish the StSS from the ordinary helical state, which also produces a stray field around the sample.
Moreover, MFM is not able to detect the relative orientation of the surface spin spirals on the opposite surfaces which is desired for experimental revealing of the result of our theoretical calculations.

The most promising experimental technique, which may allow to detect the StSS seems to be Lorentz TEM. 
Note, typically the Lorentz TEM images provide  information only about the in-plain component of the  magnetization. 
Figure~\ref{TwistAv} shows the in-plane component of the magnetization averaged over the film thickness. Two cases with stripe- and square-like pattern presented in figure~\ref{TwistAv}(a) and (c) corresponds to the spin structure shown in figure~\ref{Twist}(a) and (d), respectively.

\section{Method}\label{method}
To construct the phase diagram shown in figure~\ref{PD}(a), we performed an energy minimization for each of the states and compared their energy density, see for instance figure~\ref{PD}(b). 
A finite difference approximation was used to convert the Hamiltonian (\ref{E_tot_m}) into a function, which arguments are  the components of unit vectors defined on the nodes of finite size grid with unit cell size $\Delta x\times\Delta y\times \Delta z$. 
Such calculations have been done for certain set of reduced thicknesses, $L/L_\mathrm{D}$, defined by the geometry of the simulated domain, material parameters and applied fields varying in the range $0\le H/H_\mathrm{D} \le 1.5$.
For the function minimization we use a nonlinear conjugate gradient method implemented on NVIDIA CUDA architecture.
In order to achieve the highest performance of the code and to keep the constraint $\mathbf{n}^2=1$, we use the method of adaptive stereographic projections introduced in Ref.~\cite{Rybakov_15}.

The size of the simulated domain along each of the spatial directions is defined as $l_x=N_x \Delta x$, $l_y=N_y \Delta y$, $l_z=(N_z - 1)\Delta z$, where $N_x$, $N_y$ and $N_z$ are the number of nodes fixed to 128, 256 and 256$...$512, respectively (up to $16\times 10^{6}$ nodes in total).
For the fixed thickness, $L$, the value of $\Delta z$ is chosen such to satisfy the equality $L=(N_z-1) \Delta z$,
while $\Delta x$ and $\Delta y$ are defined self-consistently to identify such $l_x$ and $l_y$, which correspond to the lowest average energy density of the state.
We assume periodic boundary conditions in the $xy$-plane and open boundary conditions along $z$-axis at $z=0$ and $z=L$.
Thereby, the procedure of energy minimization for each point on the phase diagram and for each of the states consist of an \textit{a priori} unknown number of alternating steps: i) the direct energy minimization for given $\Delta x$, $\Delta y$, $\Delta z$ and ii) the small variation of $\Delta x$ and $\Delta y$. 
In most general case, the variation of $\Delta x$ and $\Delta y$ is assumed to be independent.
In particular, it is important for the proper energy minimization of StSS state, which is in general characterized by the modulations in both $x$- and $y$-directions, see figure~\ref{Twist}. 
For the case of helical spin spiral and assuming $\mathbf{k}\parallel x$-axis, the optimal $\Delta x$ minimizes the energy of the system when $l_x$ equals to the equilibrium period of the helix.
Because of the homogeneity of the helical state in the direction perpendicular to $\mathbf{k}$-vector,  the variation in $\Delta y$ does not affect the energy density of the solution at all, and the problem reduces to a quasi-two-dimensional one.

The procedure based on direct minimization of the functional discussed above had been applied for relatively thin films, $L\le 4L_\mathrm{D}$. 
For the thick films, we take into account the exponential decay of surface induced modulation \cite{Surf_twist1}, which becomes negligibly small on the distance from the surface $ > \!L_\mathrm{D}$.
Thereby, the average energy density of any modulated state in the extended (infinite along the $x$- and $y$-directions) thick film with $L>4L_\mathrm{D}$ with a very high precision can be approximated as
\begin{equation}
e(H,L) \!= e_\mathrm{v}(H)+\frac{2e_\mathrm{s}(H)}{L}, 
\label{thick-film}
\end{equation} 
%check energy density per unit volume
where $e_\mathrm{v}$ is the average volume energy density for a particular state, which is assumed to be homogeneous along the applied field and $e_\mathrm{s}$ is the average surface energy density of the twisted state. 
The $e_\mathrm{v}=\mathcal{E}_\mathrm{v}/V$, where $\mathcal{E}_\mathrm{v}$ is the total energy calculated on the domain with $L=4L_\mathrm{D}$ in the absence of free surfaces meaning periodic boundary conditions in all three spatial directions; $V$ is the volume of the simulated domain.
Then, the total energy $\mathcal{E}^\prime_\mathrm{v}$ has been calculated on the same domain with open boundary condition at $z=0$ and $z=L$.
The surface energy density $e_\mathrm{s}$ has been identified from the difference $\mathcal{E}_\mathrm{v}-\mathcal{E}^\prime_\mathrm{v}$.   
For each phase with surface induced modulations, the value of $e_\mathrm{v}$ and $e_\mathrm{s}$  has to  be calculated ones for each fixed value of $H$ and then the energy density in whole range of thicknesses $L\ge 4L_\mathrm{D}$ can be found according to Eq.~(\ref{thick-film}). 

In the range of thicknesses $4L_\mathrm{D}<L<8L_\mathrm{D}$ we performed the calculations of the phase transitions with both approaches: i) direct energy minimization in whole volume of the layer and ii) according to Eq.~(\ref{thick-film}). For this range of thicknesses we found identical results with both approaches.  

In conclusion, one has to mentioned that the finding of the global energy minima for the functional (\ref{E_tot_m}) is by no means a straight-forward task due to the following reasons.
Because of the complex energy landscape of 3D model with large energy barriers between the different states, the practically used algorithms of minimization provide only the local energy minimum close to the initial state, which in general may not correspond to the global minimum. 
To define the energetically dominant state one has to calculate and compare the energies of all competing phases. When the set of such tested states is incomplete it is impossible to identify the phase transitions properly. For instance when the earlier unknown StSS state is ignored the final phase diagram is incorrect \cite{Leonov_15arXiv}.     

\section*{Conclusions} 
We have presented the complete phase diagram for a film of isotropic chiral magnets, which allowed us to identify the range of existence of the equilibrium skyrmion lattice in applied magnetic fields and varying thicknesses.
In particular, we found the critical thickness of the film above which the skyrmions never appear as the ground state of the system.
We predict the existence of a new modulated state of \textit{stacked spin spirals}, which occupies a wide range of the phase diagram for thin films as well as for bulk crystals.
Such a state represents the coexistence of the conical spiral and surface induced spirals localized near the surface with finite penetration depth.
We are confident that the presence of such states as well as the validity of the phase diagram itself can be confirmed with various experimental techniques.

\section*{Acknowledgments}
The research of F.N.R. and A.B.B. was carried out within the state assignment of FASO of Russia (theme Quantum No. 01201463332).
The work of F.N.R. was supported in part by RFBR (project No. 14-02-31012).

%%%\section*{References} 

\end{document}